# A 2ª Lei da Termodinâmica na formulação da Lei de Hooke

Rodrigo de Abreu
Centro de Electrodinâmica e Departamento de Física
do IST

**Abstract**

Hooke's Law is analyzed taking the Second Law of Thermodynamics into consideration. It is shown that the energy of a spring isn't always $½ k x^2$ - this value is actually the energy of the spring plus the energy of the atmosphere that surrounds it. On a quasi-isentropic aproximation in which the dynamic force is approximately the same as the static force, $k$ is not a constant. Only by considering an ideal spring where $k$ is constant and only on a static regime, can we say the energy is $½ k x^2$. For any dynamic regime of such an ideal spring, the energy of the spring is a function of its temperature and therefore is different from $½ k x^2$. If any given mass is moving attached to an ideal spring, the movement will eventually cease even if such movement occurs in vacuum. The dynamic force is not the static force $kx$. The Second Law of Thermodynamics is essential in order to formulate Hooke's Law.

**Resumo**

Analisa-se, a partir da 2ª Lei da Termodinâmica, a Lei de Hooke. Mostra-se que numa transformação reversível, isentrópica, a energia de uma "mola" nem sempre é dada por $V = ½ k x^2$ que é a energia da "mola" mais a energia do ar da atmosfera onde se considera que a mola está imersa. Num regime dinâmico, quase-isentrópico, em que a força dinâmica é aproximada pela força estática, com generalidade $k$ não é constante. A energia só é $V = ½ k x^2$ para uma "mola ideal", em que $k$ é constante, ao longo de uma isentrópica, num regime estático - uma isentrópica é uma isotérmica. Num regime dinâmico, no caso limite de uma "mola ideal", a energia é função da temperatura e não pode ser expressa por $V = ½ k x^2$. Um movimento oscilatório, harmónico, linear sem atrito corresponde a uma trajectória quase-isentrópica e, portanto, uma massa em movimento, submetida à acção de uma "mola ideal", tende para o repouso, mesmo que não exista ar a perturbar-lhe o movimento. Tal significa que a força de restituição da mola não é a força estática $k\,x$. A 2ª Lei da Termodinâmica é fundamental na formulação da Lei de Hooke.

## Introdução

A distinção entre força dinâmica e força estática permite estabelecer uma relação entre a 2ª Lei de Newton e a 2ª Lei da Termodinâmica. Uma "mola" (um material elástico) submetida a uma força exterior, sofre uma variação de comprimento. Verifica-se, como é bem conhecido, uma proporcionalidade entre esta força e o comprimento adquirido, atingido que seja o equilíbrio na presença do ar da atmosfera à temperatura $T_0$. Esta é a lei de Hooke. Podemos matemáticamente exprimi-la por

$$F = k\,x. \qquad (1)$$

em que *F* é o módulo da força que actua sobre a "mola", igual ao módulo da força exercida pela "mola", num ponto de equilíbrio, em que o comprimento da mola não evolui no tempo. *k* é a constante da "mola" e *x* é o aumento de comprimento devido a *F*. Note-se que é possível generalizar esta conceptualização para diversas situações, por exemplo (no electromagnetismo) para um condensador, descrito matematicamente de forma análoga [1]. Usamos a palavra "mola", tendo em atenção esta generalidade. Do cálculo do trabalho elementar desta força *F*, *dW*, em torno de *x*, *dx*, variação elementar do comprimento da "mola", define-se a energia potencial $V=1/2\ k\ x^2$. De facto o princípio de conservação de energia impõe que o trabalho da força exterior num regime estático (que é igual em módulo ao trabalho da força de restituição num regime estático, reversível) tem de ser igual à variação de uma função *V*

$$dW = F\,dx = dV \qquad (2)$$

em que *V* é uma função da posição *x*. Com generalidade, como se irá demonstrar, mesmo para um regime estático, a energia da "mola" não pode ser dada por $V=1/2\ k\ x^2$. Numa mola ideal em que *k* é constante a energia da mola só aproximadamente é dada por $V=1/2\ k\ x^2$, num regime dinâmico, em que *x* evolui no tempo. A energia só é rigorosamente $V=1/2\ k\ x^2$, para uma mola ideal num regime estático, numa isentrópica, dado que para uma mola ideal num regime estático a temperatura não varia. Num regime dinâmico a força de restituição da "mola" não é igual à força estática, a força de restituição para o mesmo valor de *x*. Esta força dinâmica é assimétrica sendo maior ou menor que a força estática conforme o movimento se esteja a fazer contra ou a favor da força de restituição. Esta força dinâmica origina uma tendência para o equilíbrio com aumento de entropia [2]. Num regime dinâmico há variação de temperatura e a esta variação de temperatura corresponde uma parcela de energia que se adiciona à parcela de energia potencial da mola ideal $V=1/2\ k\ x^2$.

De (2), integrando, temos que V, a menos de uma constante é

$$V = \tfrac{1}{2} k\,x^2. \qquad (3)$$

*F* é igual a *k x* em pontos de equilíbrio. Para um dado *dx* se *F = k x*, verifica-se (2). Admitindo que a "mola" está imersa na atmosfera, á temperatura $T_0$, o princípio de conservação de energia, expresso em (2), apenas implica que V satisfaça (3). *Com generalidade V é a energia da "mola" e da atmosfera. Como iremos demonstrar, mesmo para uma mola ideal só num regime estático, isentrópico, é que a energia da mola é dada por $1/2\ k\ x^2$*. Se a "mola" estiver isolada da atmosfera, é de admitir que deixem de se verificar as condições de validade da Lei de Hooke e, consequentemente, as condições de validade de (3). Como vamos seguidamente demonstrar, assim é. Consideremos uma "mola" (uma barra [3]), tal que

$$F = aT^2(L - L_0) \qquad (4)$$

em que *L* é o comprimento à temperatura *T* quando a força é *F* e $L_0$ é o comprimento quando a força for nula e a temperatura $T_0$. *F* só é proporcional a $x = L - L_0$ se a temperatura for $T_0$.

# 1. Cálculo da energia

Vamos determinar a energia em função da temperatura e do comprimento. Para calcularmos a energia é necessário conhecer a capacidade calorífica para um dado comprimento. Admitamos que [3]

$$C_{L_0} = \left(\frac{\partial U}{\partial T}\right)_{L=L_0} = bT. \qquad (5)$$

Dado que

$$dU = FdL + TdS \qquad (6)$$

comecemos por determinar a entropia $S$. De $S = S(T,L)$ temos

$$dS = \left(\frac{\partial S}{\partial T}\right)_L dT + \left(\frac{\partial S}{\partial L}\right)_T dL \qquad (7)$$

em que,

$$\left(\frac{\partial S}{\partial T}\right)_L = \frac{C_L}{T}, \qquad (8)$$

e

$$\left(\frac{\partial S}{\partial L}\right)_T = -\left(\frac{\partial F}{\partial T}\right)_L = -2aT(L - L_0). \qquad (9)$$

De (7), (8), (9) e (5) temos

$$(dS)_{L=L_0} = \frac{C_{L_0}}{T} dT = bdT \qquad (10)$$

e

$$(dS)_{T=T} = -2aT(L - L_0)dL. \qquad (11)$$

De (10) e (11), integrando, temos

$$S(T, L_0) = S(T_0, L_0) + b(T - T_0) \qquad (12)$$

e

$$S(T, L) = S(T, L_0) - 2aT\frac{(L - L_0)^2}{2} \qquad (13)$$

ou seja

$$S(T,L) = S(T_0, L_0) + b(T - T_0) - aT(L - L_0)^2. \quad (14)$$

Diferenciando (14) e substituindo em (6) obtém-se por integração

$$U(T,L) = U(T_0, L_0) + b\left(\frac{T^2}{2} - \frac{T_0^2}{2}\right) - \frac{1}{2}aT^2(L - L_0)^2. \quad (15)$$

De (15) concluímos que a variação de energia, numa transformação em que a temperatura final é $T_0$ (a temperatura da atmosfera) é

$$\Delta U = U(T_0, L) - U(T_0, L_0) = -\frac{1}{2}aT_0^2(L - L_0)^2, \quad (16)$$

isto é

$$\Delta U = -\frac{1}{2}kx^2, \quad (17)$$

e a variação de energia da barra é

$$\Delta U = -\Delta V. \quad (18)$$

Saliente-se que esta transformação não tem de ser necessàriamente reversível (ver 3.) e que a variação de energia da barra é o simétrico de *1/2 k x²*.

## 2. Cálculo da energia numa transformação isentrópica

Se aumentarmos o comprimento da mola progressivamente, por acréscimos infinitesimais da força exterior ($dS+dS_0=0$, $dS_0$ é a variação elementar da entropia da atmosfera e dado a atmosfera ser "infinita" a sua temperatura $T_0$ não varia, $dT = dT_0 = 0$) temos que o trabalho da força exterior (igual ao trabalho da força interior), $W$, é

$$W = \Delta V = \frac{1}{2}kx^2 = \Delta U + \Delta U_0 = -\frac{1}{2}kx^2 + \Delta U_0, \quad (19)$$

em que $U$ é a energia da mola e $U_0$ é a energia da atmosfera. Note-se que a variação de energia da mola é neste caso simétrica da variação que geralmente é considerada, a variação de energia da mola é *-1/2 k x²*. A soma das variações das energias da mola e da atmosfera é que é dada, a menos de uma constante, por *V=1/2 k x²*.

De (19) conclui-se que a <u>variação de energia da atmosfera</u> na transformação em que a soma das variações das entropias da mola e da atmosfera não variam é

$$\Delta U_0 = kx^2. \quad (20)$$

Calculemos seguidamente a variação de energia numa transformação isentrópica, agora apenas da mola, $dS=0$. De (14) temos

$$b(T - T_0) = aT(L - L_0)^2 \qquad (21)$$

ou

$$T = \frac{bT_0}{b - a(L - L_0)^2}. \qquad (22)$$

(22) define a transformação isentrópica (da "mola") em função de $T$ e $L$. Substituindo (22) em (15), temos

$$U(T,L) = U(T_0, L_0) + \frac{1}{2} a \frac{(L - L_0)^2}{b - a(L - L_0)^2} bT_0^2 \qquad (23)$$

e tendo em atenção (22),

$$U(T,L) = U(T_0, L_0) + \frac{1}{2} aTT_0(L - L_0)^2. \qquad (24)$$

A energia numa isentrópica, a menos de uma constante, é dada por

$$U(T,L) = \frac{1}{2} k(T) x^2, \qquad (25)$$

em que

$$k(T) = aTT_0. \qquad (26)$$

Num regime dinâmico, numa quase-isentrópica, temos que entre dois pontos em que $\Delta T$ é aproximadamente zero, $k$ é aproximadamente constante, mas entre o ponto de equilíbrio inicial e final (ver 3.) esta aproximação não pode ser feita.

Se admitirmos que

$$F = aT^n(L - L_0), \qquad (27)$$

temos para $n = 0$ uma mola ideal, em que $k$ não depende da temperatura. Neste caso limite e admitindo que $C_L$ é constante, temos, de (7),

$$TdS = C_L dT \qquad (28)$$

isto é,

$$S(T,L) = S(T_0, L_0) + C_L \ln \frac{T}{T_0}, \qquad (29)$$

e, de (6)

$$U(T,L) = U(T_0, L_0) + C_L(T - T_0) + \frac{1}{2}k(L - L_0)^2. \qquad (30)$$

Numa transformação isentrópica, da mola isolada do ambiente $dS=0$, e de (29) $dT=0$. De (30), temos, a menos de uma constante,

$$U(T,L) = V = \frac{1}{2}kx^2. \qquad (31)$$

Numa quase-isentrópica, (31) verifica-se aproximadamente. As equações da mecânica (num sentido restrito) têm implícitas esta aproximação. Sem esta restrição, num regime dinâmico a energia (a menos de uma contante) é dada por

$$U(T,L) = C_L(T - T_0) + \frac{1}{2}kx^2. \qquad (32)$$

Deste modo temos que num regime dinâmico a temperatura de uma mola ideal varia e tal implica necessáriamente "dissipação" de energia para o ambiente, que pode apenas ser constituído por um banho de fotões à temperatura $T_0$. Nesta transformação a soma das variações de entropias da mola e do ambiente é positiva. Por este facto quando uma mola ideal oscila tende para o equilíbrio mesmo que o exterior seja vácuo (não exista ar, dado que o radiamento está sempre presente). Quando a temperatura da mola voltar a ser $T_0$ a energia da mola volta a ser $1/2\ k\ x^2$. Mas durante as oscilações a energia da mola não pode ser dada por $V=1/2\ k\ x^2$, ao contrário do que geralmente é admitido. Apliquemos a uma mola ideal uma força de módulo $F_c$ constante, dado (por exemplo) se ter pendurado uma massa de peso $F_c$. De facto a força $F_c$ é aplicada à massa de peso $F_c$. Mas o trabalho desta força entre os dois pontos de equilíbrio em que a massa está em repouso é igual ao trabalho da força dinâmica (variável no tempo) que actua sobre a mola, dado a variação de energia cinética da massa $m$ ser zero. A energia dissipada na atmosfera, após a mola atingir o equilíbrio, é dada por ( $x$ é igual a $F_c/k$)

$$F_c x - \frac{1}{2}kx^2 = F_c \frac{F_c}{k} - \frac{1}{2}k\frac{F_c^2}{k^2} = \frac{1}{2}\frac{F_c^2}{k}. \qquad (33)$$

O facto da mola regressar á temperatura $T_0$ imposta pelo ambiente faz com que, por vezes, se pense que o trabalho da força exterior é igual à variação de energia da mola. De facto metade desse trabalho é dissipado no ambiente. Só para uma mola ideal e numa transformação isentrópica é que a energia da mola é dada por $V=1/2\ k\ x^2$ e a energia dissipada é nula. Tal significa necessáriamente que a força exercida pela mola num regime dinâmico não pode ter módulo $kx$ ao contrário do que geralmente é admitido.

# 3. Determinação das variações de energia e de entropia numa transformação irreversível

Estabeleçamos as equações que permitem determinar a variação de energia da mola entre o ponto de equilíbrio inicial e o ponto de equilibrio final após a aplicação da força constante $F_c$:

Para a "mola" que satisfaz a equação (4) e cuja energia é dada pela equação (15) as equações que permitem determinar o estado final de equilíbrio são

$$F_c x = \Delta U$$
$$F_c = aT^2 x \qquad (34)$$

Estas equações têm como incógnitas $T$ e $x$ e podem fácilmente ser resolvidas.

As equações equivalentes para a mola ideal são

$$F_c x = \Delta U = C_L(T - T_0) + \frac{1}{2}kx^2$$
$$F_c = kx \qquad (35)$$

cuja solução é

$$x = \frac{F_C}{k} \qquad (36)$$

e

$$T = T_0 + \frac{1}{2}\frac{F_c^2}{k}\frac{1}{C_L}. \qquad (37)$$

A variação de entropia da mola nessa transformação é

$$\Delta S = C_L \ln\left(1 + \frac{1}{2}\frac{F_c}{k}\frac{1}{C_L}\frac{1}{T_0}\right) \qquad (38)$$

Esta variação de entropia resulta do afastamento da força dinâmica da força estática [2]. Esta variação de entropia é independente da existência ou não de atrito resultante da interacção com o ar da atmosfera. Por essa razão, (o afastamento da força dinâmica da força estática, origem do atrito interno), a variação de entropia surge no interior da mola. Se se medir o amortecimento das oscilações no vácuo pode determinar-se a força dinâmica. O amortecimento das oscilações é geralmente atribuído ao atrito externo mas de facto o atrito num movimento oscilatório é tambem devido ao atrito interno.

Experimentalmente este efeito pode ser verificado através da diminuição do atrito externo, através da diminuição da densidade do ar onde a mola se encontre imersa.

Se a temperatura e a entropia da mola regressarem aos valores iniciais após troca de energia com a atmosfera a variação de entropia da atmosfera é, de (33)

$$\Delta S_0 = \frac{\frac{1}{2}\frac{F_c^2}{k}}{T_0}. \qquad (39)$$

## Conclusões

Quando se analisa o movimento de uma massa *m* submetida a uma força de restituição baseada na lei de Hooke, é geralmente admitido que a força sobre a massa m é a força estática, a força num ponto de equilíbrio. O afastamento da força dinâmica da força estática descreve um atrito interno, dá origem a uma variação de entropia. Determinou-se a energia em função da temperatura e do comprimento. Numa quase-isentrópica a temperatura varia. Numa isotérmica a variação de energia da mola e do ambiente é ½ *k x²* - a variação de energia da mola, com generalidade não pode ser ½ *k x²*, embora, numa quase-isentrópica, possa ser dada aproximadamente por ½ *k x²*. Numa mola ideal, só numa isentrópica é que a energia é dada por ½ *k x²*. Só neste caso ideal é que uma transformação de um sistema finito (a mola), é simultâneamente isentrópica e isotérmica. Só como aproximação é que se verifica um movimento oscilatório harmónico linear sem atrito. Uma massa *m* a oscilar submetida à acção de uma mola (isolada do ambiente) tende para o repouso devido ao atrito interno – a força de restituição num regime dinâmico não pode ser *k x*. A força dinâmica pode ser experimentalmente determinada medindo-se o amortecimento no vácuo.

## Referências